\documentclass[preprint,12pt]{elsarticle}

\usepackage{amssymb}
\usepackage{lineno}

\journal{Nuclear Physics B}

\begin{document}

\begin{frontmatter}



\title{Small-$x$ Physics with the ALICE experiment at the CERN-LHC}


\author{Tapan K. Nayak {\it for the ALICE Collaboration}}

\address{Variable Energy Cyclotron Centre, Kolkata - 700064, India}

\begin{abstract}
High energy \mbox{p-p}, \mbox{p-Pb} and \mbox{Pb-Pb} 
collisions at the CERN-LHC offer unprecedented 
opportunities for studying wide variety of physics at small Bjorken-$x$.
Here we discuss capabilities of the ALICE experiment at the CERN-LHC for probing small-$x$ QCD physics.
A new forward electromagnetic calorimeter is being proposed as an
ALICE upgrade to explore the small-$x$ region in more detail.

\end{abstract}


\end{frontmatter}


% \linenumbers

\section{Introduction}
%\label{}

The general formulation of the theory of strong interactions, Quantum Chromodynamics (QCD), 
is based on the principle of Bjorken scaling which suggests that experimentally observed hadrons behave as 
collections of point-like constituents (such as quarks) when probed at high energies. The dimensionless
scaling variable,
$x=Q^2/2M\nu$, has been introduced in electron-nucleon deep inelastic scattering (DIS), where $Q^2$=$-q^2$ 
is the squared 4-momentum-transfer through the exchanged virtual photon to the target nucleon, 
$\nu=q.p/M$ is the energy loss by the electron, $M$ is the target nucleon mass. 
Recent high precision DIS data from H1 and ZEUS experiments at the HERA electron-proton collider confirm the
scaling behavior over a very wide kinematic range of the variables ($x,Q^2$).
DIS experiments have shown that at high $Q^2$ the number of quark-antiquark pairs with small-$x$ goes up. 
At small-$x$ the parton distribution functions (PDF), which give the momentum
distribution of partons in a nucleon, 
are not properly determined and have large uncertainties because of
lack of experimental data. 
The experimentally observed data from HERA \cite{hera} shows that 
the gluon density of the PDF of the proton grows as $xG(x,Q^2) \propto x^{-\lambda(Q^2)}$,
where the exponent $\lambda$ is observed to rise logarithmically with $Q^2$.
At small enough $x$, a 
saturation region occurs, where the non-linear effects of gluon-gluon fusion due to the high
gluon density becomes important. 
With increasing energy and higher number of participating
nucleons in going from eA to pA and AA collisions, one expects to enter to the 
novel region of gluon saturation.
Moreover, the PDFs within a nuclei
are not known as a nucleus cannot be treated as simple superposition of protons and neutrons. 
Thus, experiments with proton and ion collisions at the
CERN Large Hadron Collider (LHC) will be able to probe an unprecedented range of $x$-values and provide important
input for theoretical calculations.

\begin{figure}
\begin{center}
\includegraphics[width=12.5cm]{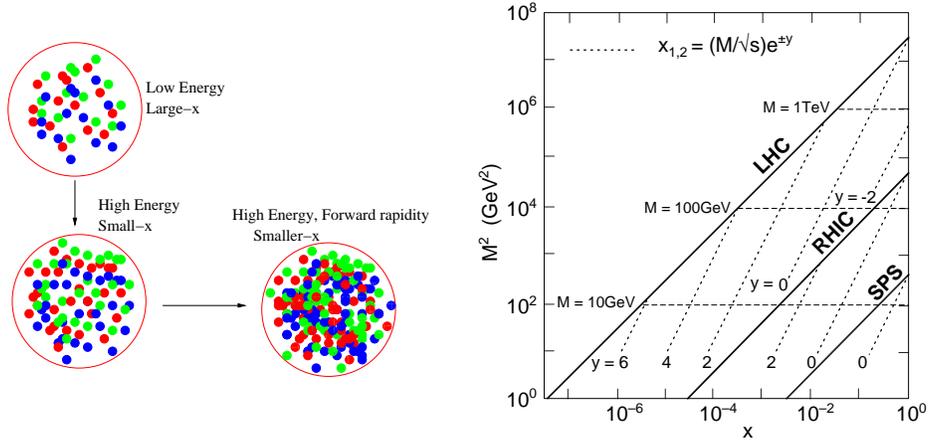}
\caption{\label{fig:xvalues} (Colour online) The left panel gives a schematic diagram of getting to the
small-$x$ saturation regime at higher energy and forward rapidity. The right panel gives 
the range of Bjorken-$x$ values accessible at 
heavy-ion collisions at the top SPS, RHIC and LHC energies \cite{alice-ppr-1}. The lines of constant
rapidity are shown.}
\end{center}
\end{figure}

Small Bjorken-$x$ (expressed as, $x=p_{\rm T}/\sqrt{(S)} \cdot e^{-y} \sim p_{\rm T}/\sqrt{(S)} \cdot e^{-\eta}$, 
where $y$ and $\eta$ are
rapidity and pseudo-rapidity, respectively) values
can be accessed by going to higher center-of-mass energy and higher rapidity as depicted in the left panel 
of Fig.~\ref{fig:xvalues}. The right panel of Fig.~\ref{fig:xvalues} gives the accessible range of Bjorken-$x$
values and $M^2$ relevant for particle production in nucleus-nucleus collisions at the top
SPS, RHIC and LHC energies~\cite{alice-ppr-1,alice-ppr-2}.
The study of small-$x$ regime, especially at forward rapidities, will be
most appropriate for getting to know the early stage of nuclear collision.
The ALICE experiment~\cite{alice-ppr-1,alice-ppr-2,alice-jinst} in the present setup will probe a 
continuous range of $x$ below 10$^{-4}$. A new proposal of installing a forward electromagnetic
calorimeter at larger rapidity (up to $\eta$=5) is being discussed which will reduce the $x$-value to less than
10$^{-5}$. Thus, ALICE experiment will be able to access
a novel regime where initial state effects like Color Glass Condensates (CGC) (for a review see \cite{cgc}) 
type of phenomenon can be studied in great detail.

The Large Hadron Collider (LHC) at CERN is designed to deliver colliding 
\mbox{p-p} beams 
at center-of-mass energies of 14~TeV and \mbox{Pb-Pb}
beams at 5.5~A TeV. LHC is also capable to provide light ion collisions such as \mbox{Ar-Ar} and 
as well as asymmetric collisions like p-Pb.
Table~\ref{tab:three} gives the center-of-mass energy and 
expected luminosity at LHC for some typical collision systems~\cite{antinori}. Data from 
\mbox{p-p} collisions will be quite useful for test of pQCD (perturbative QCD) as well as act as baseline
measurement for heavy-ions. Collisions with \mbox{p-Pb} will probe nuclear PDFs and will be useful to
disentangle initial and final state effects. The \mbox{Pb-Pb} collisions probe the hot and dense medium.
A combination of data at all these colliding systems as well as collision centralities will provide a large
window on the rich phenomenology of high-density PDFs, throw light on shadowing, gluon saturation as well as
CGC. Thus the collisions at LHC's
unprecedented energies offer outstanding opportunities for access to new physics.

\begin{table}[hbt]
\begin{center}
\begin{tabular}{||c|c|c|c||}
\hline\hline
System       & $\sqrt{s_{\rm NN}}$(TeV)     & L$_0$(cm$^{-2}$s$^{-1}$) & $\sigma_{\rm geom}$(b) \\ \hline \hline
\mbox{p-p}                    &  7/10/14    &   10$^{34}$($\sim$10$^{30}$ for ALICE)          &   ~0.1        \\
\mbox{Pb-Pb}                  &   2.75/5.5  &   10$^{27}$        &   7.7          \\
\mbox{p-Pb}                   &   8.8       &   10$^{29}$             &   1.9          \\
\mbox{Ar-Ar}                  &   6.3       &   10$^{29}$             &   2.7          \\
\hline \hline
\end{tabular}
\label{tab:three}
\caption{System, collision energies and peak luminosities expected at the CERN-LHC.}
\end{center}
\end{table}

\section{Observables to probe small-$x$ QCD phenomena}

Here we mention a sample of observables which can probe small-$x$ QCD physics.

\subsection{Global Observables}
A new hadron production mechanism emerges at small-$x$ in the context of gluon saturation and CGC
picture. The hadron production is dominated by production of gluon minijets and their fragmentation into hadrons.
At RHIC, gluon saturation picture provides one of the explanations for observed particle multiplicity, which is much smaller
than previously expected. Recently, pseudorapidity distributions of charged particles at mid-rapidity
have been published by ALICE and other LHC experiments for \mbox{p-p} collisions at 0.9TeV~\cite{alice1} 
and 7TeV~\cite{alice2}. These results could be successfully described in two recent publications by
McLerran {\it et al}~\cite{mclerran} and Levin {\it et al.}~\cite{levin} in the CGC framework. 
These calculations also make predictions for higher energy as well as heavy-ions. 
Here we show one of the figures from~\cite{levin} where the energy dependence of the charged
hadron multiplicities are shown for \mbox{p-p} and central heavy-ion collisions. Data points show a compilation
of results from different experiments. The solid bands are theoretical curves, and the one with saturation model
describes the data well. Of course, for further test the saturation mechanism, more exclusive 
measurements are needed.
\begin{figure}[hbt]
\begin{center}
\includegraphics[scale=0.4]{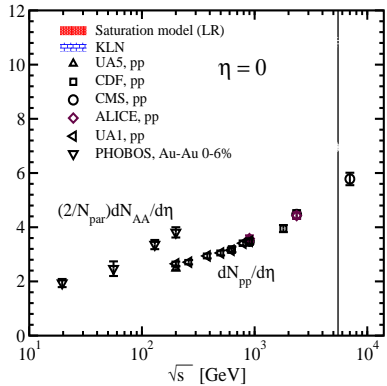}
\caption{ \label{fig:Levin} (Colour online) 
Energy dependence of the charged
hadron multiplicities for \mbox{p-p} and central heavy-ion collisions~\cite{levin}.
}
\end{center}
\end{figure}

\subsection{Nuclear Modification Factor of Hadrons}
\begin{figure}[hbt]
\begin{center}
\includegraphics[scale=0.34]{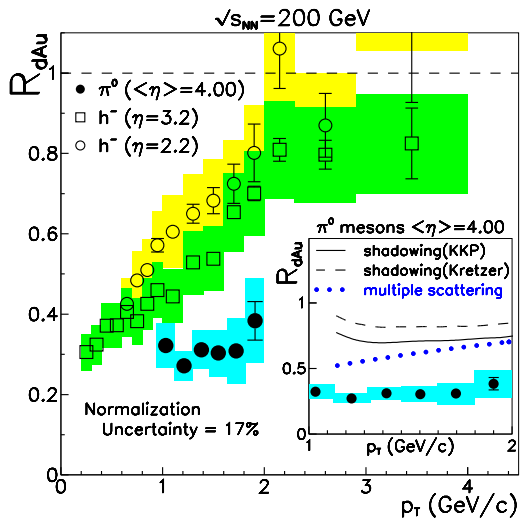}
\caption{ \label{fig:RdAu} (Colour online) Nuclear modification factor, $R_{dAu}$, as a function of $p_{\rm T}$
at forward rapidities as measured by STAR ($\pi^0$) and BRAHMS (negative hadrons) experiments.
}
\end{center}
\end{figure}

Hadronic observables directly probe the gluon distributions in the nucleon or in the nucleus. Effect of 
gluon saturation at small-$x$ can be studied by measuring particle production at high energy and large
rapidity for \mbox{p-p} and \mbox{p-A} collisions. At RHIC energies, this has been measured \cite{brahms,star}
in terms of suppression of inclusive hadron yields and broadening of azimuthal correlation related to recoil jets
from parton-parton scattering. The nuclear modification factor, $R_{dAu}$, defined as,
\begin{equation}
R_{dAu} = \frac{dN_{dAu}/dp{\rm T}} {N_{\rm coll}(dAu) dN_{pp}/dp_{\rm T}},
\end{equation}
for hadron production in \mbox{d-Au} collisions with respect to \mbox{p-p} collisions, is shown in Fig.~\ref{fig:RdAu}
for large rapidity, measured by BRAHMS~\cite{brahms} and STAR\cite{star} collaboration. Significant suppression
of the yield in \mbox{d-Au} collisions is seen, qualitatively consistent with gluon saturation models. The inset
of the figure shows that the conventional calculations including shadowing effects are not able to 
describe the observed suppression \cite{star}. Measurements at LHC will be a breakthrough in such studies.

\subsection{Heavy Quark Production}
\begin{figure}[hbt]
\begin{center}
\includegraphics[scale=0.45]{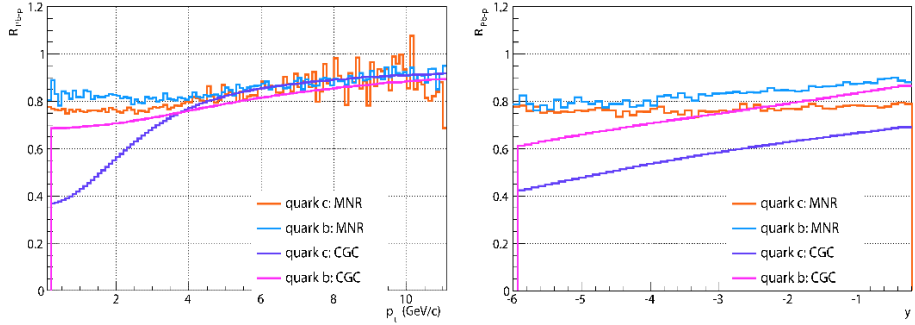}
\caption{\label{fig:RPbp} (Colour online) Nuclear modification factor ($R_{Pb-p}$) of heavy quarks 
as a function of $p_{\rm T}$ (left panel) and rapidity (right panel)~\cite{malek}.}
\end{center}
\end{figure}

At LHC energies, heavy quarks (charm and beauty) are mainly produced through gluon-gluon fusion
processes, so their production cross-sections are significantly affected by parton dynamics in the small-$x$
regime. Malek et al.~\cite{malek,charpy} have studied the heavy quark production within the CGC framework using
the ALICE simulation code. Simulations were performed for \mbox{p-p} and \mbox{p-Pb} collisions. 
The \mbox{p-p} simulations are performed with Pythia event generator tuned by MNR calculation~\cite{mnr} to
reproduce heavy quark cross-section. This uses CTEQ~\cite{cteq} parametrization of the proton PDF.
The \mbox{Pb-p} simulation were performed for two scenarios: (ii) MNR with EKS98 
parametrization~\cite{EKS98},  and (ii) within the CGC framework. In the CGC approach, the proton PDF is given by
the non-saturated CTEQ parametrization. The lead structure is described by unintegrated PDF describing the saturated nucleus.
%, defined as:
%\begin{equation}
%R_{Pb-p} = \frac{dN_{Pb-p}/dydp{\rm t}} {N_{\rm coll}dN_{p-p}/dydp_{\rm T}},
%\end{equation}
Figure~\ref{fig:RPbp} shows the nuclear modification factor as a function of $p_{\rm T}$ and y for both c-quarks and
b-quarks. One observes that at low $p_{\rm T}$, the depletion of charm is much more than that of beauty.
Thus the study of the nuclear modification factor of heavy
flavours in forward rapidity can be a suitable tool to search for the effect of gluon saturation.

\section{The ALICE Experiment}

Although the ALICE experiment \cite{alice-ppr-1,alice-ppr-2,alice-jinst} is specifically designed for heavy-ion physics, 
it is well suited for \mbox{p-p} and \mbox{p-A} collisions. 
The ALICE setup can be broadly described by three groups of detectors:
the central barrel, the forward detectors and the forward
muon spectrometer. Coverages of various detectors 
are shown in Table~\ref{tab:four} for completeness.
%\begin{figure}[hbt]
%\begin{center}
%\includegraphics[scale=0.45]{alice_sketch.eps}
%\caption{\label{fig:ALICE}The experimental setup for the ALICE experiment 
%\cite{alice-ppr-1,alice-ppr-2} 
%at the LHC. }
%\end{center}
%\end{figure}

The central barrel consists of the Inner Tracking System (ITS),
Time Projection Chamber (TPC), Transition Radiation Detector (TRD),
the Time of Flight (TOF) detector and the electromagnetic calorimeter.
The design goal is to have low material budget and low magnetic field (B$\le$0.5T)
in order to be sensitive to low-$p_T$~ particles.

The ALICE forward muon spectrometer will study the complete spectrum of heavy quarkonia 
($J/\psi$, $\psi'$, $\Upsilon$, $\Upsilon'$, $\Upsilon''$) via their decay in the $\mu^+\mu^-$ channel.
The spectrometer acceptance covers the pseudorapidity interval $2.5\le \eta \le 4$ and 
the resonances can be detected down to zero transverse momentum. The invariant mass resolution 
is of the order of 70~MeV in the $J/\psi$ region. 
\begin{table}[hbt]
\begin{center}
\begin{tabular}{||c|c|c||}
\hline\hline
Detector           & Functionality                & Acceptance ($\eta,\phi$)         \\ \hline \hline
ITS                &  vertexing, tracking,        & $\pm 2$, 360$^\circ$             \\ 
(SPD, SDD, SSD)    &  PID at low $p_{\rm T}$       &                                  \\ \hline
TPC                &   Tracking, PID              & $\pm 0.9$, full $\phi$           \\ \hline
TRD                &   Electron ID                & $\pm 0.84$, full $\phi$          \\ \hline
TOF                &   PID                        & $\pm 0.9$, full $\phi$           \\ \hline
HMPID              &   PID at high $p_{\rm T}$    & $\pm 0.6, 1.2^\circ - 360^\circ$   \\ \hline
PHOS               &   Photon spectrometer        & $\pm 0.12$, $220^\circ - 320^\circ$\\ \hline
EMCAL              &   EM Calorimeter             & $\pm 0.7$, $80^\circ - 187^\circ$    \\ \hline
ACORDE             &   Cosmic trigger             & $\pm 1.3$, $-60^\circ - 60^\circ$  \\ \hline
Muon Spectrometer  & Muon pairs                   & -2.5 to -4.0, full $\phi$          \\ \hline
PMD                &   Photon Multiplicity        & 2.3 to 4.0, full $\phi$          \\ \hline
FMD                &   Charged Multiplicity       &  -1.7 to -3.4, full $\phi$       \\ 
                   &                              &  1.7 to 5.03, full $\phi$        \\ \hline
V0                 &   Trigger                    &  -1.7 to -3.4, full $\phi$       \\ 
                   &                              &  2.8 to 5.1, full $\phi$         \\ \hline
T0                 &   Trigger, timing            &  -2.97 to -3.28, full $\phi$     \\ 
                   &                              &  4.71 to 4.92, full $\phi$       \\ \hline
ZDC(ZN and ZP)     &   Zero Degree Calorimeter    &  8.8                             \\ 
ZEM                &   EM Calorimeter             &  4.8 to 5.7, partial $\phi$      \\ \hline
Proposed Forward   &   EM Calorimeter             &  2.5 to 5.0, full  $\phi$        \\ 
Calorimeter        &                              &                                 \\ \hline
\hline
\end{tabular}
\label{tab:four}
\caption{Coverages of detector subsystems and their functionality in 
the ALICE experiment. A new forward calorimeter is being proposed to specifically
address small-$x$ physics.}
\end{center}
\end{table}

The ALICE experiment is equipped with a set of forward detectors, such as the
a Forward Multiplicity Detector (FMD), Photon Multiplicity Detector (PMD),
Zero Degree Calorimeters (ZDC),  and detectors for trigger and timing (V0, T0). 
The FMD, consisting of several rings of silicon detectors,
provide charged-particle multiplicity information.
The V0 detector provides minimum-bias triggers in pp and A--A collisions. 
The T0 detector, with two arrays of Cherenkov counters, has an excellent time resolution provides the
start time for TOF. 
The Photon Multiplicity Detector (PMD) in ALICE has been designed to measure the  multiplicity and 
spatial distribution of photons in the forward rapidity region.
The photon measurements in combination with those of charged particles 
provide vital information in terms of the limiting fragmentation,
elliptic flow of photons, and formation of disoriented chiral condensates.

\section{Proposal for a new Forward Calorimeter}
A new proposal for a Forward (Electromagnetic) Calorimeter is being considered as 
an upgrade of the ALICE experiment to explore the new small-$x$ regime of QCD. The detector is expected to cover a range
of $2.5<\eta<5.0 $ with full azimuth. It should be
capable of measuring photon energies at least up to 200GeV/c and have good
$\gamma$-$\pi^0$ discrimination capabilities. The detector will be placed at a distance of 360cm from the interaction
point on the opposite side of the muon arm. It will be a highly segmented, sampling calorimeter of about
22 radiation length. It will incorporate
tungsten as the absorber material with silicon readouts. 
Detailed simulation to optimize the detector geometry and to understand the longitudinal and lateral shower profiles,
two gamma separation, $\pi^0$ reconstruction, etc. is in progress. This calorimeter will be the major detector
for probing small-$x$ physics in ALICE.

\section{Summary}

We have presented the capabilities of the ALICE experiment at the CERN-LHC for several observables sensitive to small-$x$
values. The LHC is capable to accelerate and collide \mbox{p-p}, \mbox{p-Pb} and \mbox{Pb-Pb}. With a combination of these
data sets, one can get a detailed knowledge about new phenomena such as gluon saturation and color glass condensate. 
We have shown how the global observables, shadowing of hadrons and heavy flavour
production are affected at small-$x$. 
A new electromagnetic calorimeter at forward rapidity is proposed to specifically address the small-$x$ physics.

%% \aA new proposal for a Forward (Electromagnetic) Calorimeter (FoCAL) is being considered as an upgrade of the ALICE experiment to explore this new regime of QCD.ppendix


%% References


%\bibliographystyle{elsarticle-num}



\end{document}